\documentclass[conference]{IEEEtran}
\IEEEoverridecommandlockouts
\pdfoutput=1
\usepackage{cite}
\usepackage{pdfpages}
\usepackage{amsmath,amssymb,amsfonts}
\usepackage{algorithmic}
\usepackage{graphicx}
\usepackage{textcomp}
\usepackage{xcolor}
\usepackage{ulem}
\usepackage{changes}
\def\BibTeX{{\rm B\kern-.05em{\sc i\kern-.025em b}\kern-.08em
    T\kern-.1667em\lower.7ex\hbox{E}\kern-.125emX}}
\usepackage{flushend}
\begin{document}

\title{Redefining Information Retrieval of Structured Database via Large Language Models\\}

\author{\IEEEauthorblockN{1\textsuperscript{st} Mingzhu Wang}
\IEEEauthorblockA{\textit{School of Data Science} \\
\textit{University of Science and Technology of China}\\
\textit{ACAM and FinChina Joint Laboratory}\\
Hefei, China \\
wangmz@mail.ustc.edu.cn}
\and
\IEEEauthorblockN{2\textsuperscript{nd} Yuzhe Zhang}
\IEEEauthorblockA{\textit{School of Management} \\
\textit{University of Science and Technology of China}\\
Hefei, China \\
zyz2020@mail.ustc.edu.cn}
\and
\IEEEauthorblockN{3\textsuperscript{rd} Qihang Zhao}
\IEEEauthorblockA{\textit{Kuaishou Technology} \\
\textit{}\\
Beijing \\
zhaoqihang@kuaishou.com}
\and
\IEEEauthorblockN{4\textsuperscript{th} Junyi Yang}
\IEEEauthorblockA{\textit{School of Mathematics} \\
\textit{University of Science and Technology of China}\\
Hefei, China \\
rrrita@mail.ustc.edu.cn}
\and
\IEEEauthorblockN{5\textsuperscript{*} Hong Zhang \thanks{*Correspoding Author}}
\IEEEauthorblockA{\textit{School of Management} \\
\textit{University of Science and Technology of China}\\
Hefei, China \\
zhangh@ustc.edu.cn}
}

\maketitle

\begin{abstract}
Retrieval augmentation is critical when Language Models (LMs) exploit non-parametric knowledge related to the query through external knowledge bases before reasoning. The retrieved information is incorporated into LMs as context alongside the query, enhancing the reliability of responses towards factual questions. Prior researches in retrieval augmentation typically follow a retriever-generator paradigm. In this context, traditional retrievers encounter challenges in precisely and seamlessly extracting query-relevant information from knowledge bases. To address this issue, this paper introduces a novel retrieval augmentation framework called ChatLR that primarily employs the powerful semantic understanding ability of Large Language Models (LLMs) as retrievers to achieve precise and concise information retrieval. Additionally, we construct an LLM-based search and question answering system tailored for the financial domain by fine-tuning LLM on two tasks including Text2API and API-ID recognition. Experimental results demonstrate the effectiveness of ChatLR in addressing user queries, achieving an overall information retrieval accuracy exceeding 98.8\%.
\end{abstract}


\section{Introduction}

% \includepdf[pages={1}]{duibi_acc.pdf}

\begin{figure}[htbp]
\centerline{\includegraphics[width=\linewidth]{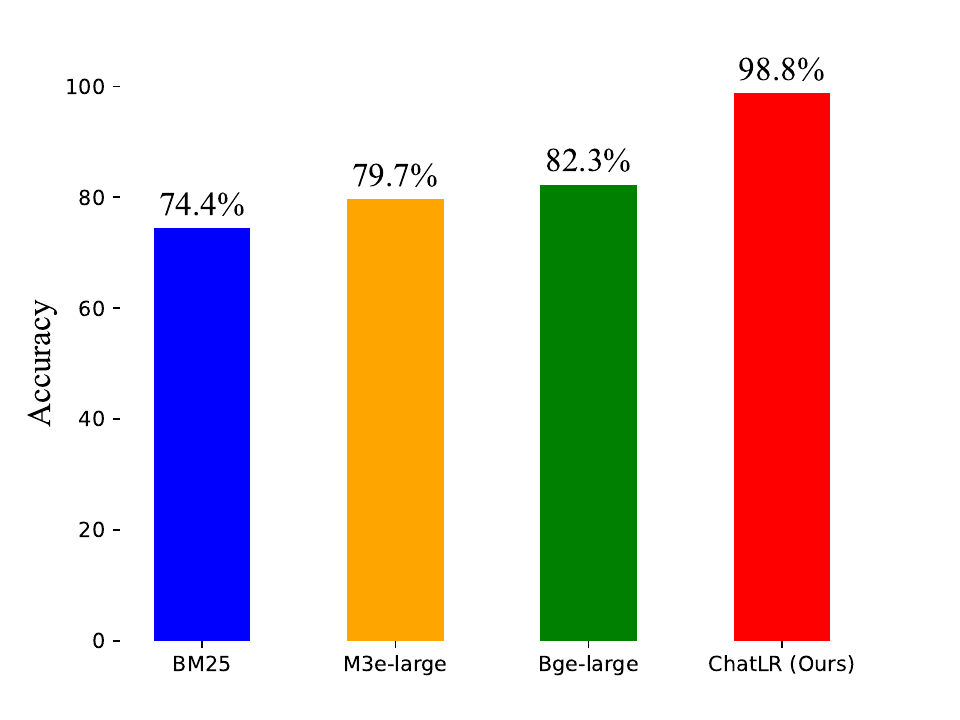}}
\caption{Performance of ChatLR (Ours).}
\label{oaacc}
\end{figure}

\begin{figure*}[htbp]
\centerline{\includegraphics[width=\linewidth]{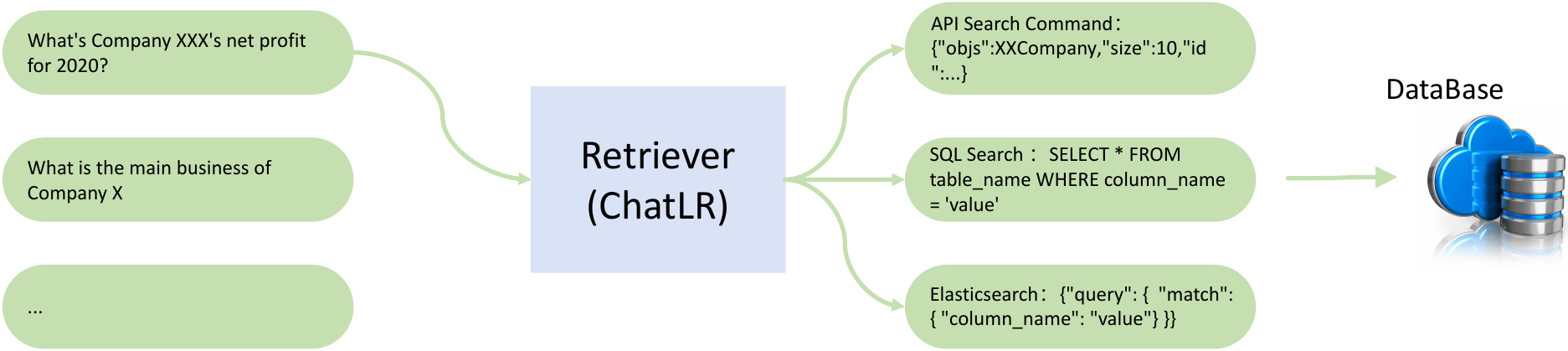}}
\caption{An example of ChatLR. ChatLR facilitates the retrieval of relevant information within a database by transforming queries (e.g., ``What’s Company XXX’s net profit for 2020?") into search command statements, such as API search commands or SQL queries.}
\label{kj}
\end{figure*}

Knowledge-intensive tasks, such as question answering and dialogue generation \cite{3,6,20}, which demand intricate memorization of factual knowledge, face challenges in achieving optimal performance directly on LLMs. This is attributed to the fact that a standalone Large Language Model (LLM) can only incorporate limited knowledge up to the point when it was trained, and it is a significant cost to update its parameters with the latest knowledge \cite{3}. Consequently, relying on implicit reasoning process of LLMs can sometimes lead to inaccuracies and complexities, hindering their application in knowledge-intensive tasks \cite{8,9,37}. Another critical issue arises from the well-known ``hallucination problem" in LLMs, where they tend to generate factually inaccurate or erroneous responses \cite{38,39,40}. This problem persists across various scenarios involving LLMs and becomes more pronounced when generating responses in vertical domains or with private data.

A promising approach to address the aforementioned challenges is Retrieval-Augmented Generation (RAG) \cite{9}. Instead of relying solely on the model's parameterized knowledge for inference, RAG enhances LMs by extracting knowledge relevant to the query from an external corpus using a retriever model \cite{4}. Subsequently, the retrieved information serves as prior knowledge, combined with the query, and is then fed into the LLMs to guide its reasoning during the generation process. In this way, the retrieved knowledge significantly aids in overcoming challenges stemming from the reliance on parameterized knowledge stored within the parameters of LLM \cite{14}. The RAG approach essentially equips the LLM with a knowledge base containing accurate and extensive knowledge. When the LLM encounters challenges in solving problems outside its expertise, it can consult this knowledge base to find relevant information before proceeding with the resolution. Moreover, the cost of updating the knowledge base is much smaller compared to retraining the LLM, emphasizing the practical advantages of timely knowledge updates.

A typical RAG framework consists of two main components: the retriever and the generator. The retriever is responsible for extracting information relevant to the query from a database, while the generator, usually a Language Model (LM), analyzes the retrieved information and the user's query to produce the final response. In previous RAG frameworks, the retriever's operation typically involves embedding external information and the query into vector, calculating the similarity between them, and selecting the most helpful information for answering the query based on vector similarities \cite{27,28}. Recent advancements have focused more on using LMs to provide supervisory signals for training the retriever. This involves the joint training of the retriever and the LM: if the LM identifies useful documents during output generation, the retriever's objective should encourage higher rankings for those documents \cite{14}. In cases where the LM is a parameter-frozen black box, supervised signals generated by the LM are utilized to optimize only the retriever to make it inclined to select information that reduces the perplexity of the text generated by the LM \cite{26}.

The information selected by the retriever plays a crucial role in the ability to respond to queries of LM. When the reference material provided to the LLM contains inaccurate content, the LM can hardly generate correct answers. Therefore, the retriever's search results should not only be highly relevant but also contain minimal noise. What we require, in essence, is a retrieval model capable of providing precise and concise information. This need is particularly pronounced when addressing questions in vertical domains such as finance and law, where accuracy and conciseness in information retrieval results are paramount. However, current information retrieval frameworks, even after fine-tuning for downstream tasks, still cannot reach satisfactory accuracy in real applications.

To address the aforementioned issue, this paper introduces a novel retrieval augmentation framework termed : LLM Retrieval with Chat (ChatLR), which leverages LLM for retrieval operations. ChatLR is specifically designed to cater to factual inquiries concerning structured databases. The integration of LLMs significantly enhances the accuracy of knowledge retrieval related to queries. Building upon the precision of data relationships stored in structured databases, our information retrieval framework achieves accurate and efficient searching of relevant information within the structured database by mapping natural language queries to precise database search commands. In the event of changes to the scenario or database, one can effortlessly generalize the framework to various vertical domains by modifying the relevant content in accordance with the ChatLR framework.

In the context of structured databases, we leverage the inherent characteristic of databases possessing structured search commands. We delineate the objective of the retriever as the generation of JSON-formatted search command statements, rather than directly retrieving data. This approach transforms the retriever's task into a sequence-to-sequence task amenable to targeted fine-tuning by LLMs. Leveraging the powerful semantic understanding and generation capabilities of LLMs, we can directly employ an LLM to generate specific commands based on the query text and sufficient information about the database search commands. In cases where the database utilizes raw SQL queries, we perform the Text2SQL task \cite{25}. This framework is adaptable to various types of queries in different scenarios. When dealing with LLMs like GPT-4 as a black-box model, a common approach is to utilize few-shot learning to stimulate the model's reasoning abilities \cite{25}, though the drawback is that the accuracy of few-shot learning may not be satisfactory. As for the case that LLMs are trainable, fine-tuning can be applied for specific tasks to achieve higher accuracy.

The contribution of this paper can be summarized as follows:
\begin{itemize}
\item We propose a novel retrieval augmentation framework named ChatLR. Utilizing ChatLR, we have engineered an intelligent system based on LLM for search and question-answering, interfacing seamlessly with structured databases. By decomposing the targets generated by queries into API-ID recognition task and Text2API task, we conducted experiments to evaluate the effectiveness of ChatLR.
\item For the tasks of API-ID recognition and Text2API, we devised a novel instruction fine-tuning data format. The data generation process was executed utilizing the GPT3.5 API. Subsequently, an extensive manual review and refinement process was undertaken to enhance the quality of the data, resulting in nearly 70,000 instances of high-quality fine-tuning data.
\item Finally, we conducted fine-tuning experiments on the open-sourced LLM named Chinese-Alpaca-33B-Pro \cite{34}. The final testing results revealed that the fine-tuned ChatLR achieved accuracies exceeding 99.9\% and 98.9\% for two specific tasks, with an overall retrieval accuracy of 98.8\%. Furthermore, it showcased superiority over other information retrieval algorithms such as BM25 \cite{43}, M3e-large \cite{42}, and Bge-large \cite{41} (see Figure \ref{oaacc}). These findings underscore the efficacy of the ChatLR framework in structured data retrieval, validating its performance against established alternatives in the field.
\end{itemize}

\section{Related Work}
\subsection{Retrieval Augmentation} \label{Retrieval Augmentation}

Incorporating retrieved information from external sources into LMs has proven to be effective in a wide range of knowledge-intensive tasks, including factual question answering, fact checking, etc. \cite{3,6,8}. Previous works encapsulated the retriever-generator paradigm, leveraging the advantage of generative models \cite{9} and neural retrievers \cite{10,8,11}. Rather than solely relying on inherent knowledge and reasoning capabilities, retrieval augmentation approaches enhance the LM by integrating a retriever component capable of sourcing knowledge from the external corpus \cite{8,9,12}. Specifically, previous retrieval augmentation methods \cite{13,14} required fine-tuning the core fractions of LM to adapt to the retriever on specific downstream tasks. With the emergence of LLMs, there have been endeavors to employ LLMs as generators for question-answering tasks under few-shot and zero-shot setting after the information retrieval process \cite{3,4,5}. Since then, fine-tuning LLMs becomes prohibitively expensive as the number of unique demands continue to increase \cite{15}. Moreover, many state-of-the-art LLMs can only be accessed via black-box APIs \cite{16,17}. These APIs enable users to submit queries and receive responses but generally do not support fine-tuning. To enhance the accuracy of LLMs in answering user queries, substantial prompt engineering or fine-tuning retrievals using small-scaled LMs is required is necessary \cite{4,5}.

\subsection{Structured Knowledge-Base Retrieval Augmentation} \label{Structured knowledge-base Retrieval Augmentation}

Retrieval augmentation techniques can be applied to both structured and unstructured knowledge bases. For retrieval from structured knowledge bases, Sun et al. \cite{21} explored open-domain question answering only from web tables without leveraging the unstructured text data. Recent studies have delved into Machine Reading Comprehension (MRC) with tables, although without a retrieval module \cite{22,19,23}. Furthermore, Chen et al. \cite{24} conducted research on open-domain question answering utilizing both tabular data and textual sources.

Retrieval Augmentation of primary unstructured databases is the process of retrieving relevant information from unstructured data, such as text and knowledge graph, according to the user queries, and then making LMs generate responses. This line of research is commonly referred to as TextQA, primarily designed for open-domain question answering tasks from textual data. Previous approaches often relied on graph neural networks for fine-tuning or adopted linearization of structured knowledge and combine it with texts \cite{7.1,7.2,7.3,7.4}. By framing user queries, structured knowledge, and outputs in the text-to-text format \cite{2}, our work aims to advance the field of retrieval augmentation for structured database.

In this paper, we focus on a structured knowledge base (SKB) that restore substantial amounts of authoritative data, often presented in tabular form. It is worth noting that, in contrast to relational databases and traditional KBs, tables can be best described as semi-structured information. In the financial field, structured tabular data can, to a certain extent, ensure data accuracy and consistency. Therefore, when performing retrieval augmentation on SKB, it is crucial to prioritize the retrieval performance. Previous studies have employed various methods, including relational chains and deep neural networks \cite{21}, beam search \cite{22}, iterative retrievers \cite{24}, among others. However, this paper primarily focuses on utilizing LLMs for retrieval augmentation for structured databases.

\begin{figure*}[htbp]
\centerline{\includegraphics[width=\linewidth]{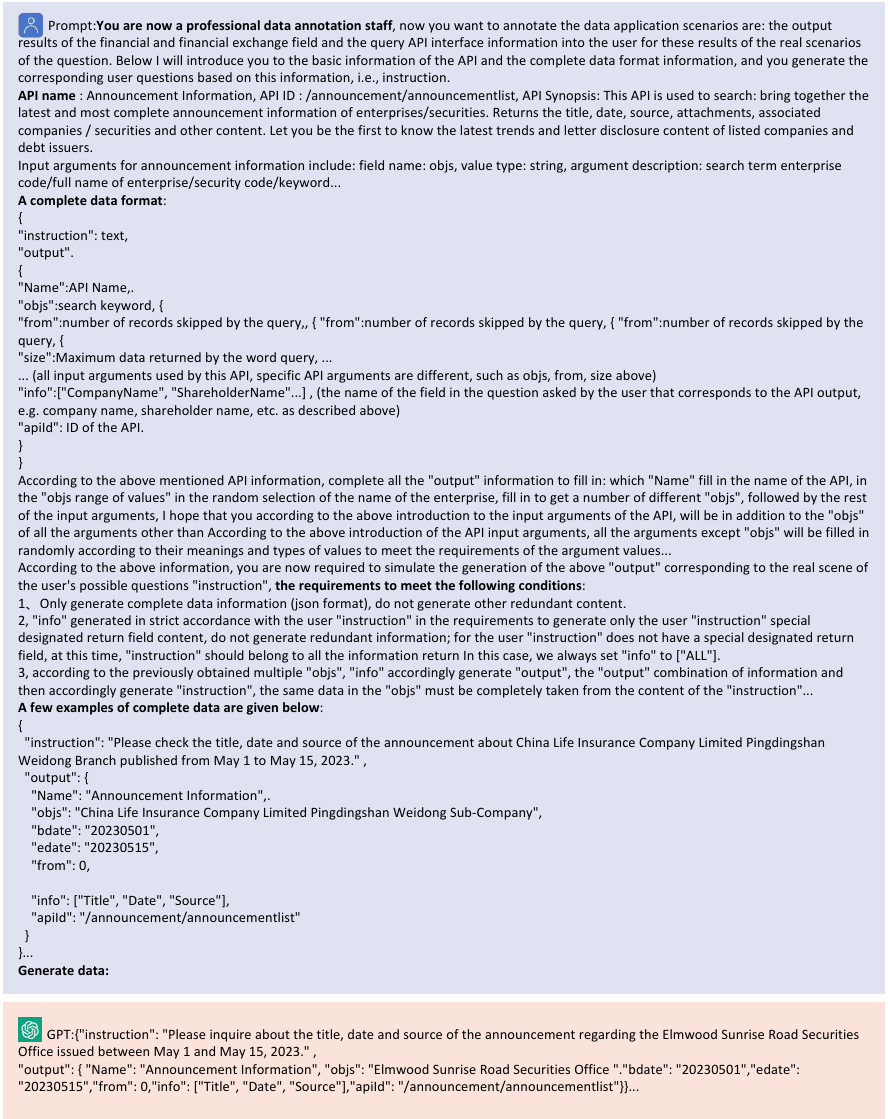}}
\caption{An example of data generation process.}
\label{sjsc}
\end{figure*}

\section{Method}

In this section, we first present our ChatLR framework in subsection \ref{The Overview Of ChatLR Framework}, followed by an introduction to the fine-tuning data generation paradigm in subsection \ref{Data Generation Paradigm}, and concluding with an overview of the fine-tuning methodology in subsection \ref{Fine-Tuning}.

\subsection{The Overview of ChatLR Framework} \label{The Overview Of ChatLR Framework}

Previous information retrieval algorithms such as BERT \cite{28} and Contriever \cite{27}, typically embed external knowledge and queries into dense vectors, and select top $k$ similar results from the external knowledge base. Generally, retrieval algorithms can be fine-tuned for downstream tasks. For instance, REPLUG \cite{26} utilizes LMs to generate supervised signals, optimizing the retrieval component to align with the LM. However, even after fine-tuning, the algorithms that rely on dense space retrieval struggle to achieve satisfactory accuracy and concise information retrieval.

In response to factual inquiries directed from structured databases, we propose a novel information retrieval framework called ChatLR that achieves both precise and refined information retrieval by mapping natural language queries to precise database search commands. As show in Figure \ref{kj}, specifically, the query interfaces for structured databases include several types: specific API commands, which provide fixed output information based on specific input arguments; SQL searches, which involve operations such as returning table data and calculations through SQL codes; full-text searches, which utilize query keywords to index the database for word matching, and so forth.

For the aforementioned retrieval commands, we can leverage LLMs to generate specific commands based on text data, such as performing tasks like Text2SQL \cite{25}. With this framework, we delegate information retrieval tasks to LLMs, capitalizing on the extensive semantic understanding capabilities of large language models. This framework can adapt to various types of questions in different scenarios. When the LLM is a black-box model like GPT-4, a common approach is to utilize few-shot learning to stimulate the model's reasoning abilities \cite{25}, though accuracy remains challenging to ensure. In the case of open-source trainable models like LLAMA \cite{35}, fine-tuning for specific tasks can be performed to achieve higher accuracy. In the following sections, we will demonstrate the effectiveness of our framework through specific task applications in particular scenarios.

\subsection{Data Generation Paradigm} \label{Data Generation Paradigm}

With the release of LLMs such as ChatGPT \cite{16} and GPT-4 \cite{17}, researchers can utilize their APIs for the sake of data generation and augmentation on domain-specific tasks with their powerful language comprehension and generation capabilities. The data used for fine-tuning are generated through the ChatGPT API with prompt constructed by concatenating information from database search interfaces with specific commands. In essence, what we require is for the GPT model to simulate questions that a user might pose in real-world scenarios regarding information in the database. These questions serve as inputs to the model, and the corresponding database search commands are annotated as model outputs.

Our data generation process closely follows the self-instruct approach \cite{29}. An example of our data generation process is shown in Figure \ref{sjsc}. Initially, we prepare a seed question file, which guides data generation via GPT-3.5 API. Additionally, we incorporate prompt engineering techniques tailored for ChatGPT. Firstly, we establish the context and role, for instance, instructing the model: ``You are currently a data annotation expert in a specific domain." Secondly, we provide sufficient reference information and offer clear explanations. This includes specific interpretations of database search command arguments, the range of argument values, required data formats, etc. Lastly, following the In-context Learning (ICL) method \cite{30,31}, we introduce several concrete data examples, enabling the model to emulate these examples and generate standard, meaningful, and accurate data. Through iterative improvements in data generation quality, we observed that summarizing the shortcomings of previously generated data and explicitly addressing these in the prompt significantly enhanced the quality of data generated after prompt updates. Ultimately, with the support of prompt engineering, we utilized the GPT-3.5 API to generate a batch of fine-tuned data that played a crucial role in subsequent experiments, aligning well with our requirements. Prior research has highlighted the significance of sample quality over quantity in fine-tuning LLMs \cite{49}. Consequently, following the completion of data generation, a substantial human effort was invested in meticulously reviewing and modifying the fine-tuning data. This rigorous review aimed to ensure the rationality of the data, including considerations such as the feasibility of generated $instructions$ in real-world scenarios and the correspondence between the content of $instructions$ and the command contents in the $output$. Through this meticulous process, a substantial volume of high-quality fine-tuning data was obtained.

\begin{figure*}[htbp]
\centerline{\includegraphics[width=\linewidth]{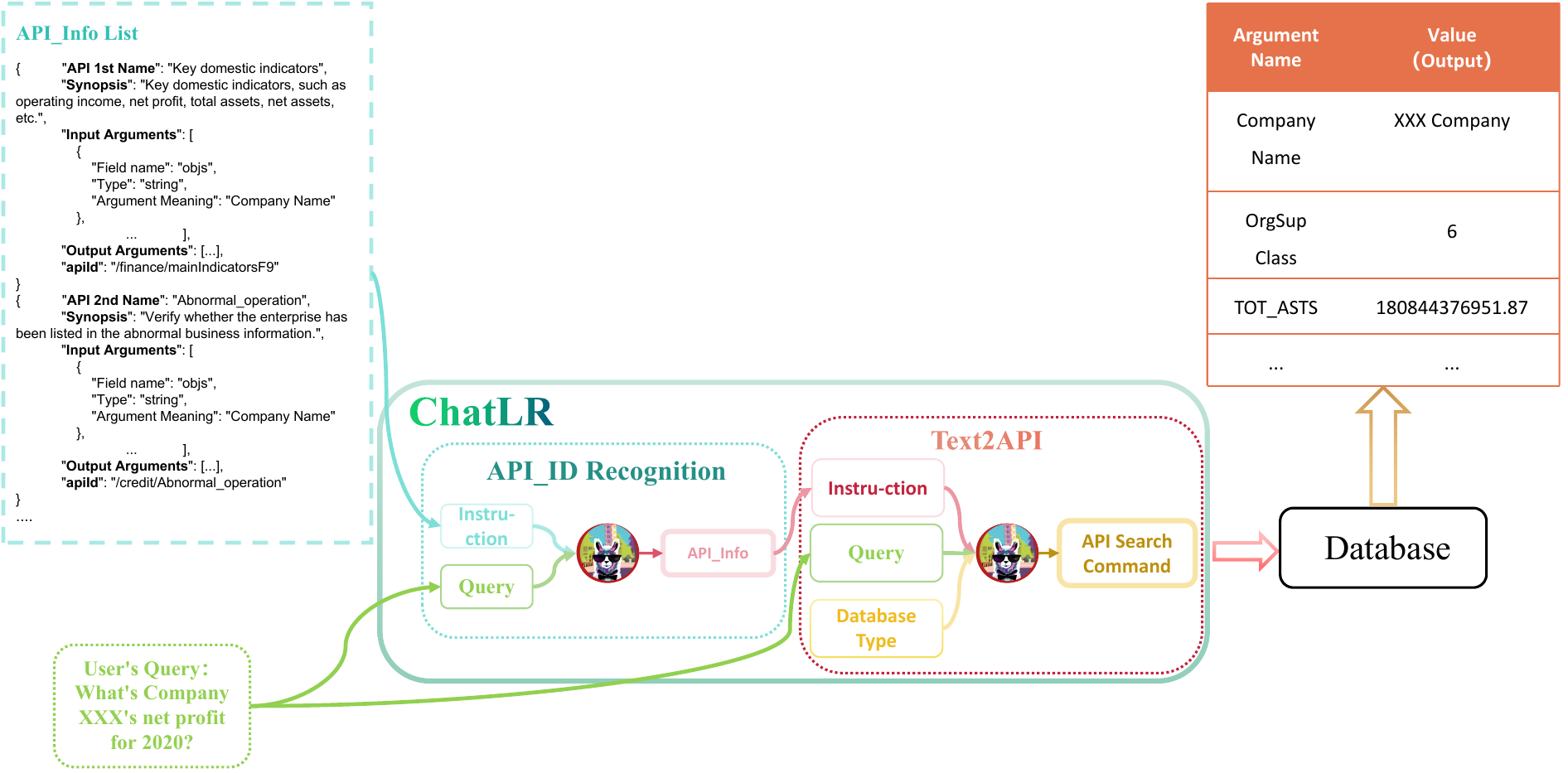}}
\caption{A Retrieval System for Structured Databases. ChatLR undertakes two distinct tasks. Initially, it engages in API-ID recognition, parsing the user's query to identify the specific API-ID. Subsequently, leveraging the API-Info list, it retrieves detailed information about the identified API, including input and output argument names. With the acquired API information, ChatLR proceeds to execute the Text2API task, generating the precise API search command statement. Finally, the search command is employed to retrieve accurate results from the database, which are then presented to the user.}
\label{chatlr}
\end{figure*}

\subsection{Retrieval System Based on LLM for Structured Databases} \label{Fine-Tuning}

To validate the effectiveness of the ChatLR retrieval framework, we integrated a specific structured database from financial domain, undertaking the entire process from data construction to fine-tuning LLM instructions. This comprehensive approach allows us to explore the practical application of the ChatLR retrieval framework for structured information retrieval in a specific financial context. Should there be alterations in the scenario or database, adapting the pertinent content in line with the ChatLR framework allows for the seamless application of the framework to different vertical domains. The database utilized specific API search commands, where users can choose specific API functions (e.g., key domestic indicators API providing functionality to retrieve specific arguments such as operating income). Inputting specific argument values yields precise result tables within the diverse functionalities provided by the API.

Figure \ref{chatlr} shows the practical application of the ChatLR framework as a retrieval system for structured databases. Specifically, ChatLR observes real-world user queries to identify key arguments. The arguments include specific API functionality required to address the query, the filling values for the API input arguments, and specific return values needed from the multitude of results provided by the API. We integrate these arguments into JSON-formatted labels, forming fine-tuning data in conjunction with user queries. 

As for the process of converting user instructions into specific API command information, it is imperative to provide the LLM with the comprehensive information of respective API as a prerequisite. However, due to the substantial number of APIs, incorporating the entire set of API information into the LLM prompt tokens at once would significantly surpass the maximum context length of the LLM. To address this challenge, inspired by the methods employed in tasks such as Text2SQL using LLMs like GPT-4 \cite{17}, we adopt the strategy of initially mapping the user's query to a specific table. Subsequently, we furnish the LLM with the information about that table, including attributes like headers, in conjunction with the query, facilitating the completion of Text2SQL reasoning \cite{32}.

To refine the fine-tuning task of mapping user queries to API command labels, we decompose it into two distinct sub-tasks within our framework: API-ID recognition and Text2API. Specifically, by first determining the particular API to reference, it becomes unnecessary to incorporate other irrelevant API information into the prompt, thereby effectively mitigating the maximum context length constraint. This approach is feasible due to the unique identifier assigned to each API, which serves as a distinguishing factor. Consequently, we leverage the API-ID recognition to precisely position the user's inquiry onto the specific API-ID capable of flawlessly resolving the search task. This method ensures the systematic and efficient functioning of our framework in handling token length constraints, offering a viable solution to the computational challenges posed by the vast API landscape.

\begin{figure*}[htbp]
\centerline{\includegraphics[width=\linewidth]{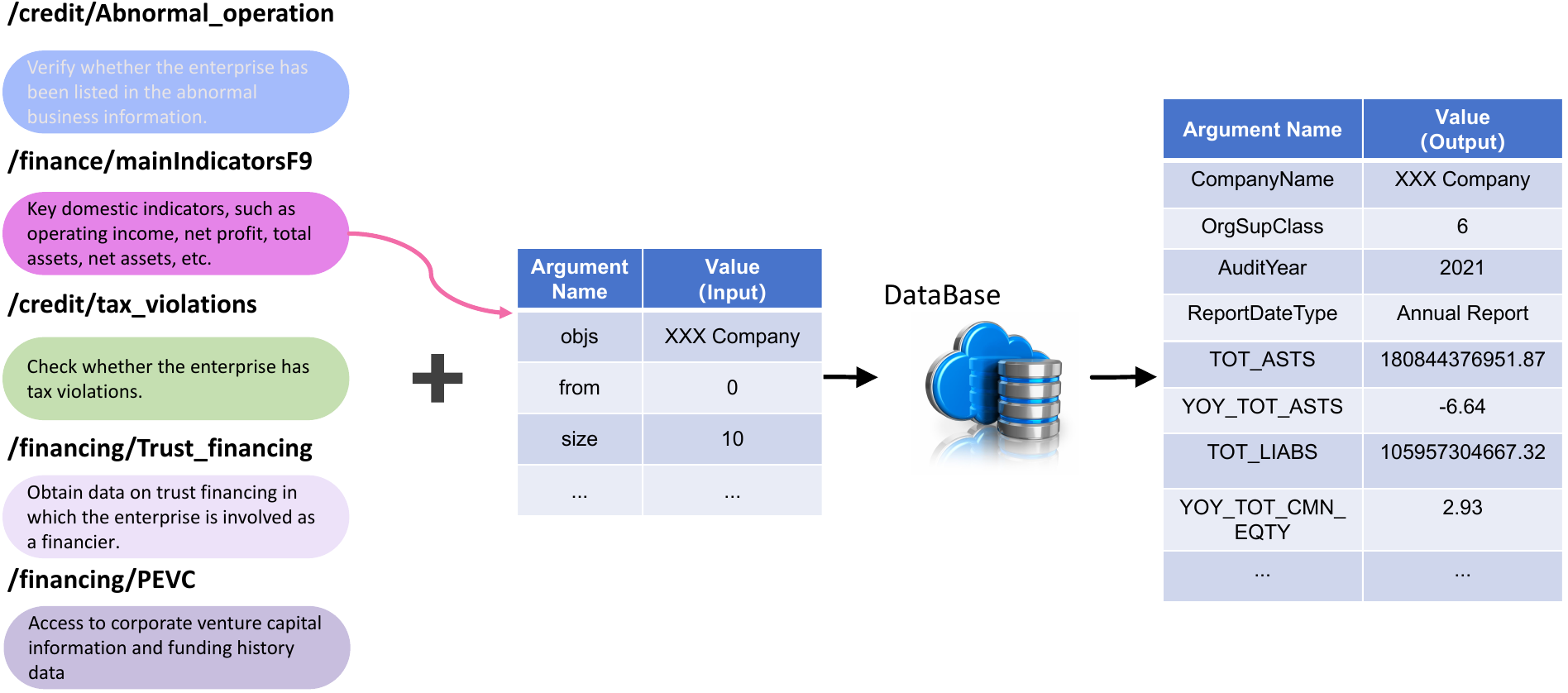}}
\caption{Structured database API search framework. Users select the appropriate API-ID based on the functional descriptions of various APIs, inputting all required argument values for the chosen API. Subsequently, a database query is performed to obtain comprehensive tabular data.}
\label{fig}
\end{figure*}

The specific procedure involves concatenating a command prompt template with specific instructions, along with all API-IDs and their corresponding Chinese names, with the user's query. The concatenated prompt is then fed into ChatLR. Subsequently, ChatLR produces a specific API-ID. Upon completing API-ID recognition, the obtained result is fed into API-info, a database containing fundamental information about all APIs as illustrated in the Figure \ref{fig}. This information encompasses ID, all required input arguments along with their value types, argument meanings, and output argument details. This information primarily serves to provide sufficient external knowledge for LLM in subsequent prompt engineering.

We extract the information of the identified API, compile it into an API-INFO prompt template, concatenate it with the user's query, and further supplement it with special command information. This composite input is then reintroduced into ChatLR. This time, due to the distinct command, the model outputs a standardized JSON-formatted API command. This command includes API-ID, API input arguments with correct values, and info (a list comprising all necessary user-desired output argument names). Finally, the corresponding argument values are input into the identified API retrieval function with the help of API command information file. This process involves querying the external database to retrieve relevant information, extracting output argument information included in info, and reorganizing it to present to users. To summarize, the framework, facilitated by LLM, serves as a bridge, transforming user natural language queries into precise search commands, resulting in accurate and refined financial data.

\section{Experiment} 

In this section, we introduce our experiments in detail, including datasets in subsection \ref{Datasets}, models in subsection \ref{Models}, implementation details in subsection \ref{Implementation Details}, and experimental results in subsection \ref{Results}.

\subsection{Datasets} \label{Datasets}

The structured database used in our experiments is an exhaustive repository encompassing financial, legal, and associated data facets. This comprehensive resource restores essential information pertaining to major companies and organizations, including financial indicators, financing details, judicial case specifics, etc. The dataset is securely stored in an SQL database maintained by FinChina Company. Nevertheless, owing to the substantial volume and extensive breadth of the data, the efficiency of SQL query statements may occasionally fall short of practical application demands. In response to this challenge, a strategic solution has been implemented by integrating 60 API query interfaces tailored to the specific type and scope of the database information. This integration allows users to rapidly and precisely retrieve required information within the expansive database through the API interface. 

As show in Figure \ref{fig}, during the information retrieval process, each API mandates particular argument values (e.g., an organization's name). Depending on the API's functionality and the input argument values, it returns a list of output arguments with accurate information. Our objective is to leverage LLMs for the retrieval of user queries related to pertinent knowledge from the database. This endeavor seeks to deliver results that are not only accurate but also refined.

According to the data generation method mentioned in subsection \ref{Data Generation Paradigm}, we generated approximately 30,000 pieces of raw data for 60 APIs. After two rounds of filtration, including program-based filtering and manual review and modification (conducted by FinChina Company), a total of 11,000 pieces of raw data were obtained. Program-based filtering ensured the correctness of the data in JSON-format and verified whether the data for specific APIs included all their input arguments and whether the output argument names were correct. Subsequent manual review was conducted to validate the correlation between the instructions generated by GPT and the content in the output. In case of discrepancies, necessary modifications and adjustments were made. This meticulous process was implemented to ensure the accuracy and reliability of the generated data.

Upon the foundation of the original data, we curated distinct fine-tuning datasets for the API-ID recognition task and the Text2API task. Our fine-tuning data are structured in the format of $\{instruction, input, output\}$. For the API-ID recognition task, the instruction is composed of specific instruction content concatenated with the ID and Chinese name information of all APIs. The input corresponds to the user's query, and the output represents the ID of a specific API. Regarding the Text2API task, the instruction is constructed by combining special instruction content with the overall information of a specific API. The input is the user's query, and the output is the complete JSON-formatted API search command. The design of the specific instruction content in the instruction for both tasks closely resembles the prompt structure in data generation. We incorporated prompt engineering techniques, including scenario and role setting, detailed information interpretation, and specific examples referencing ICL, into both instruction designs. Subsequent experimental results emphasize the importance of prompt content, even in the context of fine-tuning data, for LLMs.

\begin{table*}
\caption{Main results on text summary}
\centering
\resizebox{0.6\textwidth}{!}{
\begin{tabular}{lrrrr}
\hline
Model                 & BLEU-4 & Rouge-1 & Rouge-2 & Rouge-L \\ \hline
ChatGLM-6B            & 16.5 & 36.1  & 19.0  & 23.5 \\
ChatGLM-6B/FT         & 65.2 & 78.5  & 70.6  & 74.3  \\
BELLE-13B             & 15.3 & 38.2  & 25.2  & 20.4  \\
BELLE-13B/FT          & 62.1  & 80.3   & 72.3   & 73.7   \\
Chinese-Alpaca-13B    & 16.7 & 39.6  & 27.2  & 22.3  \\
Chinese-Alpaca-13B/FT & \textbf{66.1} & \textbf{85.8}  & \textbf{76.7}  & \textbf{78.3}  \\ \hline
\end{tabular}
}
\label{zy}

\end{table*}
\begin{table*}[htbp]
\caption{Main results on API-ID recognition and Text2API}
\centering
\resizebox{0.9\linewidth}{!}{
\begin{tabular}{llrcc}
\hline
Setting  & Method                             & Parameter & API-ID recognition & Text2API \\ \hline
Few-shot  & GPT3.5-turbo-16k                    & 175B       & 37.3               & 15.3     \\
          & Chinese-Alpaca33B-pro               & 33B        & 24.2               & 10.2     \\ 
Zero-shot & ChatGLM2-6B                         & 6B         & 10.2               & 3.4      \\
          & Chinese-Alpaca33B-pro               & 33B        & 15.6               & 7.2      \\
          & ChatGLM2-6B/ChatLR                  & 6B         & 75.3               & 74.4     \\
          & BELLE13B/ChatLR                     & 13B        & 78.1               & 77.3     \\
          & Chinese-Alpaca13B-plus/ChatLR (Ours) & 13B        & 89.1               & 87.3     \\ 
          & Chinese-Alpaca33B-pro//ChatLR (Ours) & 33B        & \textbf{99.9}      & \textbf{98.9}     \\ \hline
\end{tabular}
}
\label{mainr}
\end{table*}

In accordance with the outlined structure, each of the two tasks yielded approximately 11,000 instances of fine-tuning data. However, since our current data is entirely generated based on GPT, it does not fully consider the variety of queries that users might pose in real-world scenarios. After training the initial model, we subjected it to internal testing at FinChina Company, concurrently collecting data information. Subsequently, we supplemented the fine-tuning data with this additional information to enhance our model's performance and robustness in handling real-world scenarios.

During internal testing, we identified two significant issues in the API-ID recognition and Text2API tasks. For API-ID recognition, real-world user queries may not always align with solving them using one of the 60 APIs. Hence, we introduced negative samples where the query content is broad or meaningless. In such cases, we guided the model to output a negative label, providing users with a prompt indicating that the question lacks practical significance or is too broad, prompting them to rephrase for clarity and specificity. By learning from diverse types of negative sample labels, LLM can recognize complex instructions that are challenging to decompose, refusing to execute them. In such cases, the model prompts the user for more specific intent. For instance, when faced with a vague user query like ``Tell me some information," the LLM may output a prompt seeking more precise details: ``Please provide specific details for the information (company name, type of information, etc.) you want to inquire about."

Furthermore, the financial enterprise names in our original 11,000 training instances were populated from FinChina Company's comprehensive enterprise name list. However, in reality, users often prefer to inquire about specific enterprises using abbreviations or aliases. To address this, we added around 20,000 instances to the original data, using enterprise abbreviations as search keywords. With this addition, we now have approximate 30,000 original data instances in total, which were further expanded into two subtasks using distinct instructions. The final fine-tuning dataset for Text2API consists of 30,000 instances, and after introducing negative samples, the API-ID recognition fine-tuning dataset comprises 40,000 instances.

\subsection{Models} \label{Models}

In order to investigate the label alignment capability of LLMs after fine-tuning on a small yet high-quality dataset, we conducted an experiment in text summarization. Given that both our data corpus and usage context are in Chinese, we compared several state-of-the-art LLMs pre-trained on Chinese. We manually annotated summaries for 10,000 Chinese news articles and fine-tuned ChatGLM \cite{44.1,44.2}, BELLE \cite{45.1,45.2,45.3}, and Chinese-Alpaca \cite{34} using this labeled data. The experimental results (refer to the Table \ref{zy}) revealed that, irrespective of the model, all exhibited a high label alignment capability post fine-tuning. Among the three models, Chinese-Alpaca demonstrated the highest Rouge scores \cite{46} on the test set after fine-tuning. Consequently, we selected it as the foundational model for ChatLR. This model, an extension of the LLaMA \cite{35} architecture, incorporates an expanded Chinese vocabulary and undergoes continual pre-training with Chinese corpus, thereby enhancing its Chinese semantic understanding capability. Furthermore, the Chinese-Alpaca model fine-tuned with Chinese instruction data, significantly improving its understanding and execution capabilities in handling instructions \cite{34}. Our model design supports adaptation with other LLMs, providing versatility and flexibility in its implementation.

\subsection{Implementation Details} \label{Implementation Details}

In order to optimize training time and memory usage, we employed a highly efficient fine-tuning technique called LoRA \cite{36} for model training. Following a multitask fine-tuning approach \cite{33}, we simultaneously fine-tuned the model for both API-ID recognition and Text2API tasks on a dataset comprising a total of 70,000 instances. The fine-tuning process was executed on four A800 GPUs for a duration of 60 hours, resulting in the development of our ChatLR model.

\subsection{Results}
\label{Results}

\subsubsection{Metrics} \label{Metrics}

We use accuracy as evaluation metric for both API-ID recognition and Text2API tasks. For API-ID recognition task, a prediction is considered correct if it exactly matches the ground-truth label ID. Regarding Text2API task, a prediction is considered correct if the input and output argument information in the predicted result matches the ground-truth API search command information.

\subsubsection{Overall Performance} \label{Overall Performance}

From Table \ref{mainr}, it is evident that with the assistance of the ChatLR framework, different LLMs achieve significantly higher accuracy on the test sets of both tasks compared to their standalone counterparts after fine-tuning, especially under zero-shot setting. Notably, ChatLR even improves the performance of GPT-3.5 by 62.6\% in API-ID recognition and nearly 83.6\% in Text2API. For Text2API task, since we require the model to output complete JSON-formatted data and accurately infer all input arguments contained in the API along with their correct values in the given query context, it puts a higher demands on both the model's basic database-related knowledge and financial domain-specific knowledge. In light of this, our few-shot setting includes more detailed instruction explanations and incorporates multiple complete data output examples to compensate for the knowledge gap in financial domain expertise.

Due to the introduction of abbreviated and negative samples during fine-tuning, ChatLR can recognize more colloquial instructions, aligning better with the requirements for real-world applications. For instances where the user's intent is overly vague, ChatLR refrains from generating instructions and prompts the user to provide more detailed intent clarification. It exhibits the ability to consider the rationality of instructions akin to human experts, rather than rigidly transforming instructions. Furthermore, ChatLR proves to be efficient, as it only requires fine-tuning on a small-scaled pre-trained LM, ChatGLM2-6B, to achieve results that significantly surpass the 175B GPT-3.5. With high accuracy on both tasks, ChatLR allows for the accurate and rapid retrieval of target structured data based on API search commands.

\begin{figure}[htbp]
\centerline{\includegraphics[width=\linewidth]{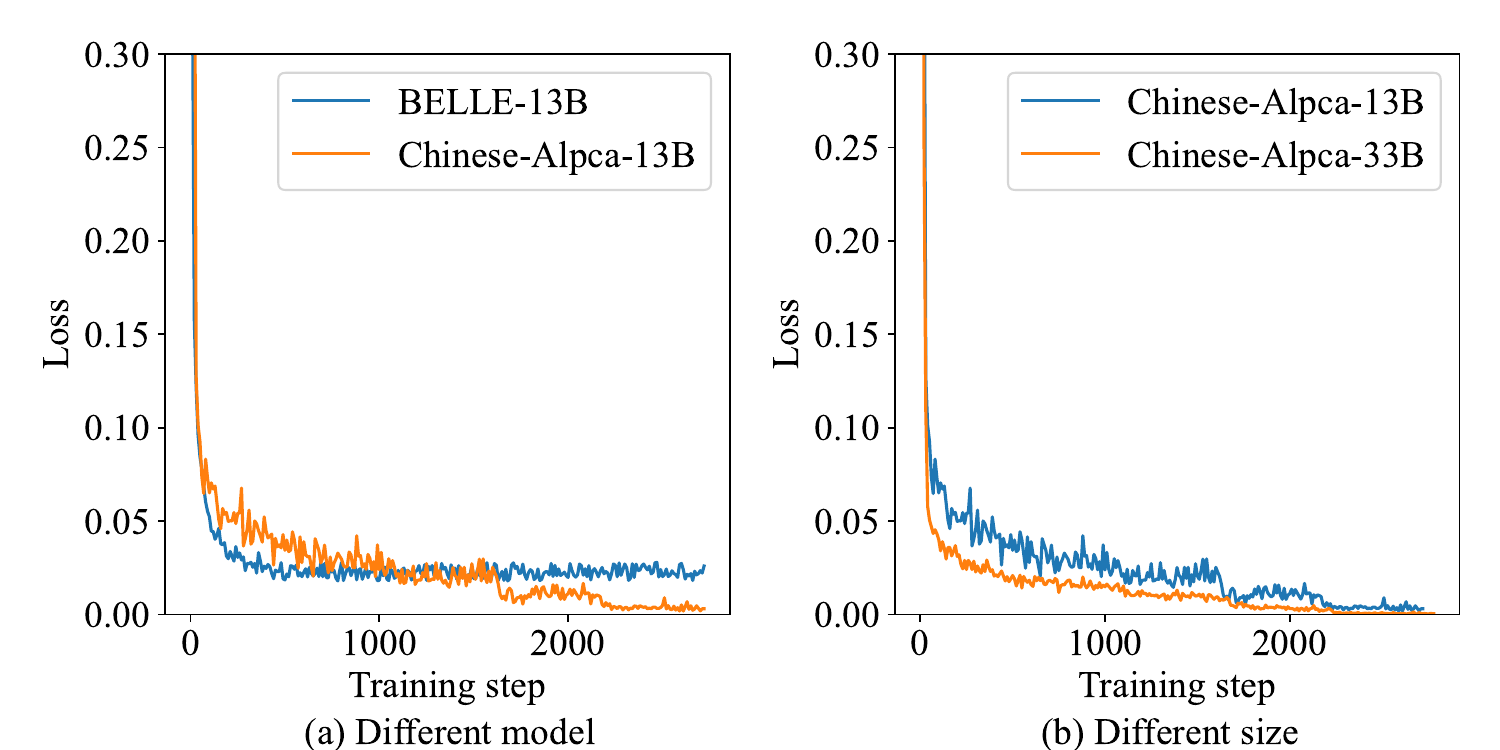}}
\caption{Training processes of ChatLR.}
\label{loss}
\end{figure}
\begin{table}[htbp]
\caption{Accuracies of various methods for information retrieval from structured databases}
\centering
\resizebox{0.25\textwidth}{!}{
\begin{tabular}{lc}
\hline
Methods      & Accuracy \\ \hline
BM25         & 74.4     \\
M3e-large    & 79.7     \\
Bge-large    & 82.3     \\
ChatLR (Ours) & \textbf{98.8}     \\ \hline
\end{tabular}
}
\label{duibi}
\end{table}

\begin{table}[htbp]
\caption{LLM-preferred Instruction Analysis.}
\centering
\resizebox{\linewidth}{!}{
\begin{tabular}{ll}
\hline
Task                    & LLM-preferred Instruction      \\ \hline
Instruction Fine-tuning & No characterization required   \\
                        & Simple and clear instructions  \\
                        & ICL not helpful                \\
Data Generation         & Requires character setup       \\
                        & Detailed cleaning instructions \\
                        & ICL helps a lot                \\ \hline
\end{tabular}
}
\label{prepro}
\end{table}

Figure \ref{loss} (a) illustrates that under the condition of equal model sizes, Chinese-Alpaca is more suitable for training the ChatLR framework. It can be seen that Chinese-Alpaca converges faster to a relatively small loss value than BELLE with continual training process. Figure \ref{loss} (b) shows that, larger foundation models lead to faster convergence. This is demonstrated by the fact that Chinese-Alpaca-33B model converges faster  to smaller loss values than its 13B counterparts with identical training steps.

\subsubsection{Comparative Experiments} \label{Comparative Experiments}

In this experiment, we conducted comparative studies on ChatLR and other baseline information retrieval models, including BM25 \cite{43}, M3e-large \cite{42}, and Bge-large \cite{41}. BM25 us a static sparse retrieval algorithm which does not need training. Both M3e-large and Bge-large are dense retrieval algorithms that are pre-trained on large unlabeled corpora and fine-tuned for specific tasks. Bge, which is inspired by the strategy of instruction tuning add \cite{45.2}, improves its general capabilities in multitask scenarios.

As demonstrated in Table \ref{duibi}, ChatLR exhibits a significantly superior performance compared to other retrieval algorithms, achieving an overall accuracy of 98.8\%. This high precision in structured data information retrieval establishes a robust foundation for various knowledge-intensive question-answering domains.

\begin{figure}[htbp]
\centerline{\includegraphics[width=\linewidth]{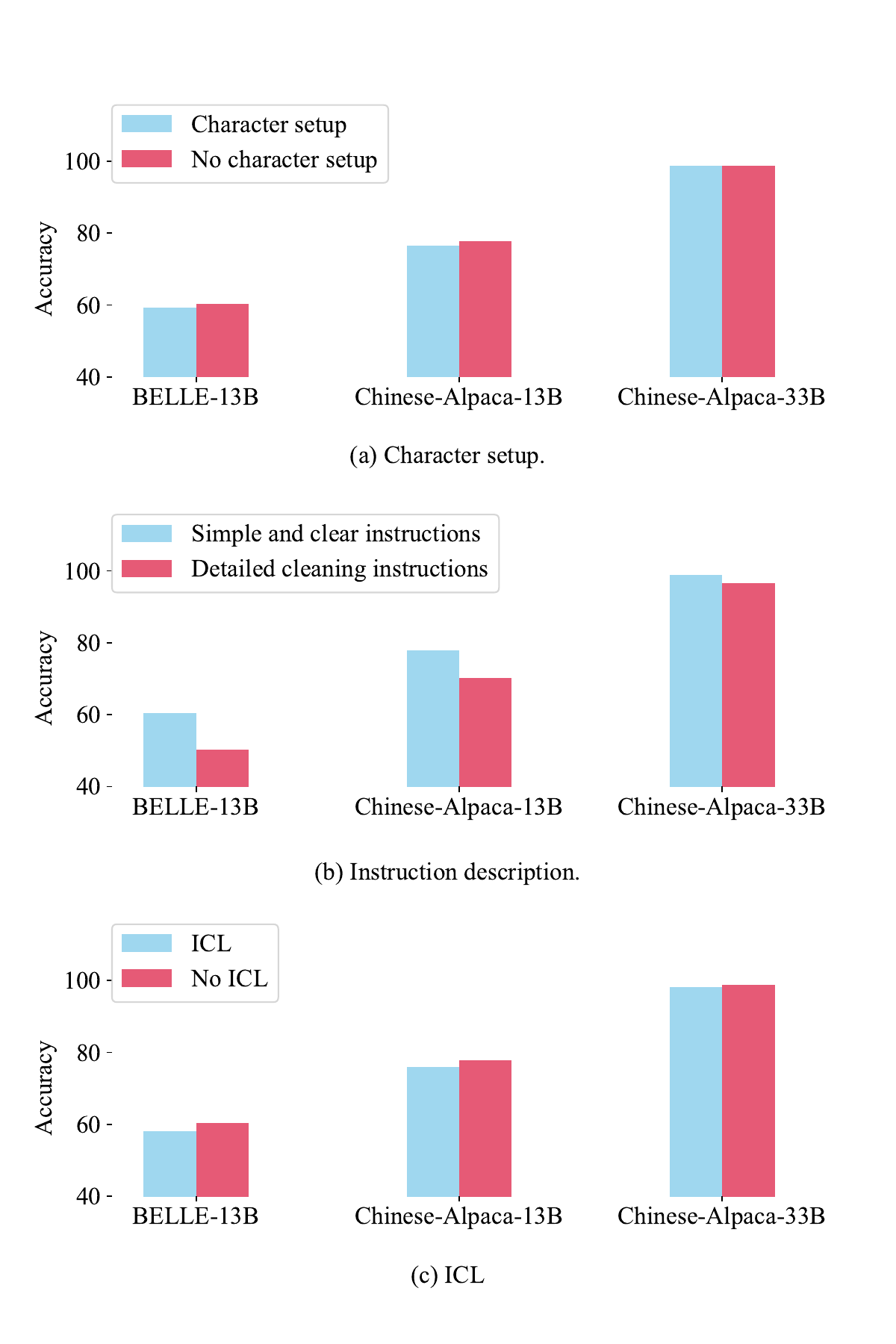}}
\caption{Accuracies of BELLE-13B, Chinese-Alpaca-13B, and Chinese-Alpaca-33B under three different instruction settings.}
\label{pre_inst}
\end{figure}

\begin{figure}[htbp]
\centerline{\includegraphics[width=\linewidth]{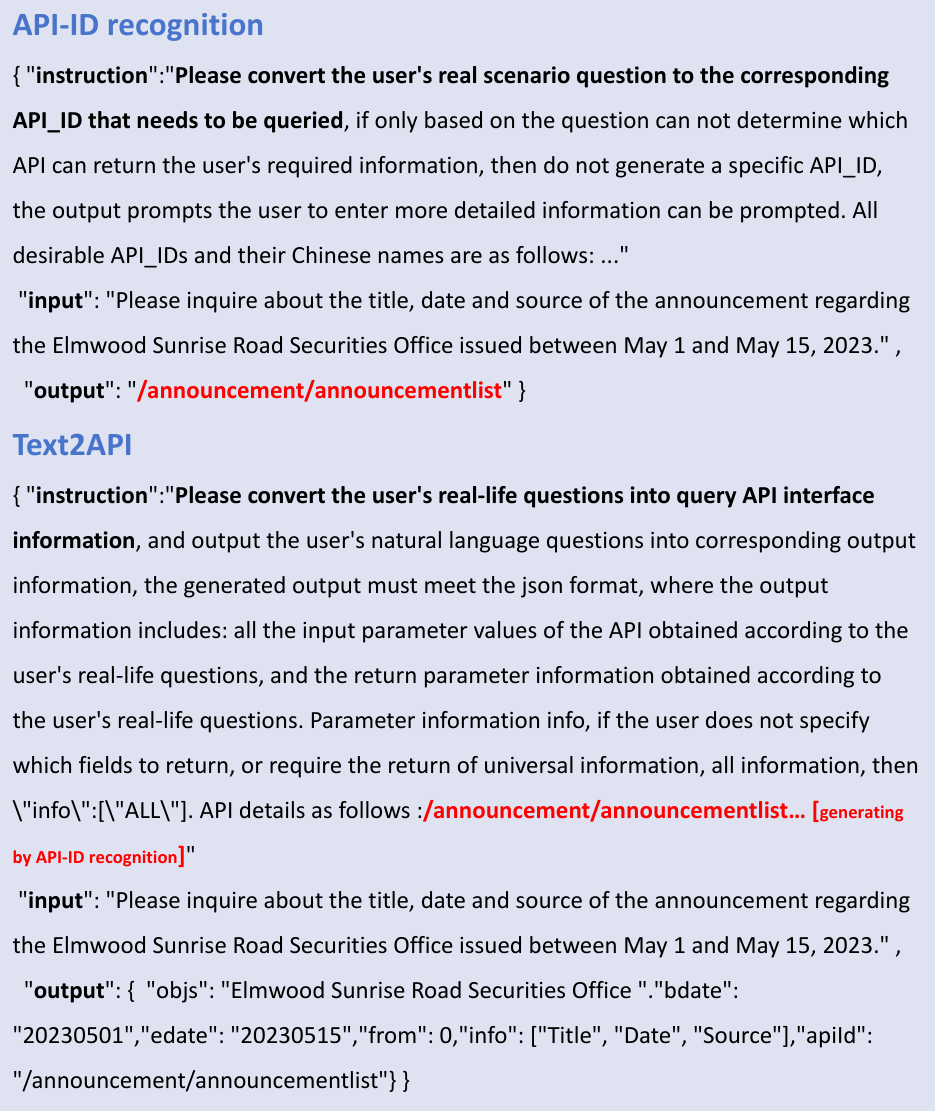}}
\caption{Two instructions for fine-tuning ChatLR.}
\label{if}
\end{figure}

\subsubsection{Analysis of LLM-Preferred Instruction} \label{Analysis of LLM-preferred instruction}

In the context of both data generation and LLM instruction fine-tuning, similar prompt content is employed. Specifically, in the inference instruction and instruction fine-tuning data, prompts are strategically used to stimulate the learning capabilities of LLM. Through extensive experiments, we note that the impact of identical prompt engineering techniques varies significantly between the inference and fine-tuning processes (see Table \ref{prepro}). We conducted multiple experiments by adjusting character setup, the level of detail in instruction content, and whether to add ICL examples separately during the fine-tuning of instructions, as illustrated in Figure \ref{pre_inst}. The experiments were performed on three foundation models: BELLE-13B, Chinese-Alpaca-13B, and Chinese-Alpaca-33B. Conclusions regarding the fine-tuning accuracy based on different fine-tuning models and parameter scales for the three instruction settings were drawn from the observations. Primarily, role-setting techniques, universally applied to LLMs, prove inconsequential during fine-tuning. Removing role-setting statements from the instructions in fine-tuning data has a negligible effect on model performance. Fine-tuning instructions necessitate clarity and directness, emphasizing simplicity. As shown in Figure \ref{pre_inst} (b), compared to the other two aspects (Figure \ref{pre_inst} (a) and (c)), this factor has a significant impact on final accuracy. For LLMs, concise instructions imply a reduced token consumption. Conversely, during data generation, comprehensive prompts are required to elaborate on how the model should assist in the task. Additionally, inspired by the principles of In-context Learning (ICL) \cite{30}, we leveraged previously accumulated high-quality data to create specific data reference examples, significantly enhancing the quality of our generated data. In contrast, the inclusion of examples in fine-tuning data instructions did not yield a notable improvement in the model's training and inference processes. Additionally, as the model scale increases, the impact of adjustments to instruction settings on the results diminishes. In other words, the larger the model scale is, the higher the robustness of this framework can achieve. For reference, we expose the specific instructions used for fine-tuning ChatLR in the Figure \ref{if}, and the instructions for data generation can be seen in the Figure \ref{sjsc}.

\begin{figure}[htbp]
\centerline{\includegraphics[width=\linewidth]{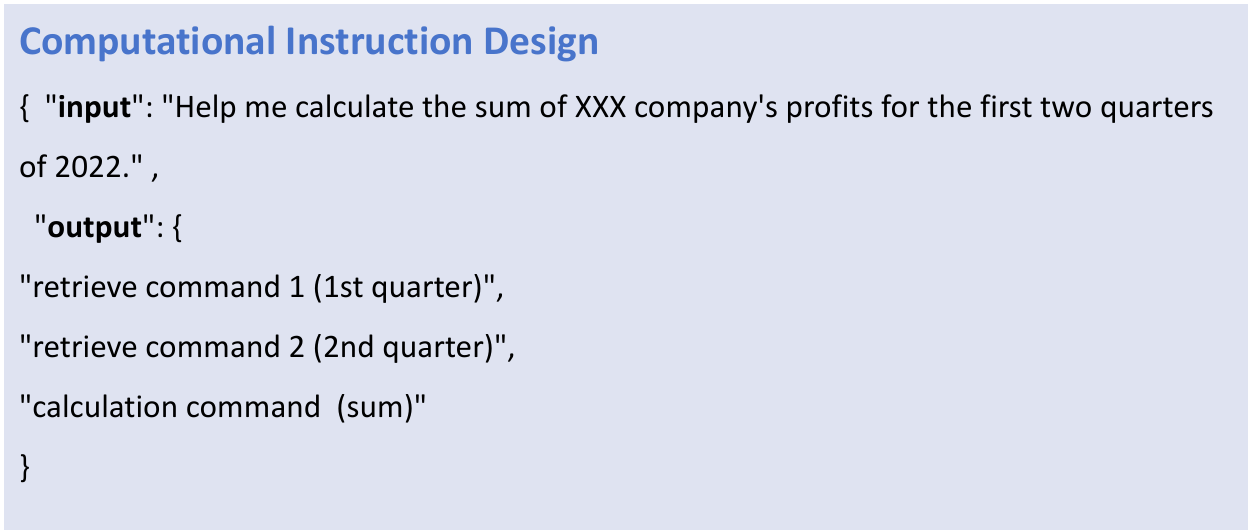}}
\caption{An Example of Computational Instruction Design.}
\label{jszl}
\end{figure}

\section{Conclusion and Future Work}

This work redefines the information retrieval framework for structured databases by incorporating LLMs and suggests fine-tuning the LLM to enhance retrieval accuracy. LLMs play a more indispensable role within the ChatLR framework, instead of a pure generator in previous RAG paradigm. Leveraging its formidable semantic comprehension capabilities, LLM can construct the mapping between queries and search commands. Consequently, it can generate precise answer utilizing the interaction between factual user queries and structured database. The effectiveness of the proposed framework has demonstrated its effectiveness on a real-world information retrieval task using domain-specific structured databases. The extensibility of this framework can be verified by moderate modification to accommodate to various scenarios and databases. Such modifications adhere to the prescribed structure, facilitating seamless generalization across diverse vertical domains. This not only substantiates the versatility of our proposed framework, but also underscores its practical utility in real-world industrial applications.

Our research predominantly focuses on optimizing queries for structured databases. However, the user queries often encompass complex intentions and requirements, such as computing maximum budgets and sorting by annual profits, which involve multi-stage mathematical reasoning steps. LLMs encounter challenges in numerical reasoning, leading to computational errors and illusions \cite{47}. As depicted in Figure \ref{if}, our proposed framework incorporates a Json-formatted output design, ensuring precise identification of commands across modules. An extended concept arising from this design involves the incorporation of calculation and statistical operation units in the output. Leveraging the natural language understanding ability of LLMs, commands related to computations and sorting are identified and outputted (as illustrated in Figure \ref{jszl}). Consequently, our approach relieves the LLM from direct operational and computational tasks, presenting a series of pre-packaged function commands for LLM invocation. By comprehending data information and user inputs, LLM generates a command sequence for execution by the backend system, thereby producing accurate computational outcomes following structured data queries, sorting, and related operations. It is noteworthy that our framework exclusively addresses the information retrieval module within the RAG framework. This exclusivity stems from the fact that the databases we utilized are structured, which obviates the need for LM to summarize outputs. For unstructured RAG with diverse data, the prevailing practice involves information retrieval through vector searches, culminating in LM summarization of outputs \cite{3,4}. In this context, LLM has the potential to optimize other RAG modules by means of its semantic understanding ability. For instance, LLM may serve as a bridge from queries to embedding keywords in retrieval from vector space. In essence, addressing complex queries often requires more than simplistic keywords to retrieve precise contextual information from databases. LLM, through query analysis, can deduce comprehensive keyword information necessary for the retriever, resulting in more accurate outcomes when matching against knowledge repositories \cite{48}.

Currently, we are endeavoring to realize the two aforementioned concepts. In future work, we aim to broaden the categories of compatible database types. Additionally, we intend to investigate an integrative approach to unify the tasks of querying from structured and unstructured databases. This goal is driven by the objective of optimizing the utilization of external database knowledge in the realm of factual knowledge-based questioning.

\normalem
\bibliographystyle{IEEEtran}
\bibliography{ChatLR}

\end{document}